\documentstyle[preprint,aps,pra]{revtex}

\tightenlines

\def\Hion{\mbox{H}_2\kern-.1em{}^+}
\def\Dion{\mbox{D}_2\kern-.1em{}^+}
\def\HDion{\mbox{HD}^+}
\def\Dmol{\mbox{D}_2}
\def\Hmol{\mbox{H}_2}

\begin{document}
\draft
\preprint{CfA No. 4476}
\title{
The hyperfine structure of the hydrogen molecular ion\thanks{Invited
talk given at the Inaugural Conference of the Asia Pacific Center for
Theoretical Physics, Seoul, Korea, June 4--10, 1996, to
be published by World Scientific.}}

\author{J.~F. Babb}
\address{
Institute for Theoretical Atomic and Molecular Physics,\\
Harvard-Smithsonian Center for Astrophysics,\\
60 Garden Street, Cambridge, Massachusetts 02138
}

\maketitle

\begin{abstract}
Theoretical investigations of the hyperfine structure of the hydrogen
molecular ion (one electron and two protons) are discussed.  The
nuclear spin-rotation interaction has been found to be of the same
sign as in the hydrogen molecule and the hyperfine transition
frequencies can be accurately predicted.  With measurements of the
hyperfine structure of the deuterium molecular ion (or of $\HDion$) it
should be possible to obtain a value of the deuteron quadrupole moment
that could be compared with the values obtained from the deuterium
molecule and from nuclear theory.
\end{abstract}

\newpage
\section*{INTRODUCTION}
The calculation of hyperfine structure (hfs) frequencies presents a
substantial challenge to theory, in part because the experimental
spectroscopic data available are so precise. For the hydrogen atom in
the $1s$ state, with electron spin $S=\case{1}{2}$ and nuclear spin
$I=\case{1}{2}$, the transition frequency between the $F=0$ and $F=1$
states, where ${\bf F}={\bf I}+{\bf S}$, is known to be
$1\,420.405\,751\,766\,7 (9)\; \mbox{MHz}$~\cite{Maser}, in
comparison to the most complete theoretical result of $1\,420.405\,1
(8) \mbox{MHz}$~\cite{Sap95}, which includes reduced mass, QED, and
hadronic (mainly proton recoil and size) effects. Indeed, a simple
``Fermi'' theory of the hfs (see, for example,~\cite{Ramsey-book},
p.~74) with a fixed proton and the electron and proton $g$-factors of,
respectively, $2.002\,319$ and $5.585\,694$, gives about $1\,422\;
\mbox{MHz}$ and with the correction due to the reduced mass of the
proton-electron system it yields $1\,420.5\; \mbox{MHz}$. Further
inclusion of the known QED corrections yields a value of about
$1\,420.452\; \mbox{MHz}$~\cite{SapYen90} and an additional $\sim
-45$~kHz and the theoretical error comes from hadronic
effects~\cite{Sap95}. Needless to say, the difficulty of the
calculations increases substantially at each stage~\cite{JauRoh76}.

The ``hydrogen atom'' of molecular physics is the hydrogen molecular
ion $\Hion$ consisting of two protons bound by a single electron.
From a theoretical point of view this molecule offers a good starting
point for investigations  of molecular hfs since the
electronic wave function can be calculated to high accuracy.
Surprisingly, there have been only a few experiments on the hfs of
$\Hion$~\cite{RicJefDeh68,Jef69,FuHesLun92}, one experiment on
$\HDion$~\cite{WinRufLam76,HD+-note} , and no experiments on $\Dion$.

With the deuterated ions it should be possible to determine the
electric quadrupole moment of the deuteron, once the hfs transition
energies are measured, and a further discussion of the prospects of using
$\Dion$ to do this will be given below.

\section*{THEORY of H${}_2{}^+$ HFS}
As with the hydrogen atom, the hfs in $\Hion$  arises primarily from
interactions between 
the electron spin and nuclear spin,
but there are also smaller magnetic effects arising from the electron
and nuclear spins interacting with the magnetic field of
the rotating nucleus.

The hfs energy  can be described by the effective spin Hamiltonian 
\begin{equation}
\label{ham}
H_{\rm hfs} = b{\bf I}\cdot{\bf S} + 
     c I_z S_z+ d {\bf S}\cdot{\bf N}  + f{\bf I}\cdot{\bf N}  ,
\end{equation}
where $\bf I$, $\bf S$, and $\bf N$ are, respectively, the total
nuclear spin, electron spin, and rotational angular momenta, and $b$,
$c$, $d$, and $f$ are the coupling constants. Transition frequencies
are obtained by diagonalizing the matrix elements of (\ref{ham}) with
constants calculated from theoretical expressions or if given measured
frequencies the constants can be determined by fitting using matrix
elements of (\ref{ham}); usually the $b_{\beta S}$~\cite{FroFol52}
angular momentum coupling scheme is used in which the intermediate
vector ${\bf F}_2 = {\bf I}+{\bf S}$ is formed, and then the total
${\bf F}={\bf F}_2+{\bf N}$.

Early theoretical studies of $\Hion$ hfs were motivated by a need for
predictions of transition frequencies of the $N=1$ rotational level of
the $v=0$ vibrational state for astrophysical searches (it is now
believed that there is little prospect for detection,
cf.~\cite{VarSan93}), because no experimental data were available.
The researchers calculated coupling constants and transition
frequencies using (\ref{ham}) with $f=0$.  Transition frequencies were
estimated by Burke~\cite{Bur60} using calculations of $b$, $c$, and
$d$, and by Mizushima~\cite{Miz60} who used values of $b$ and $c$
calculated by Stephens and Auffray~\cite{SteAuf59}.  A significant
advance in the theory was the derivation by Dalgarno, Patterson, and
Somerville~\cite{DalPatSom60} of the $b$, $c$, and $d$ terms of
(\ref{ham}) from the Dirac eq for $\Hion$ by a nonrelativistic
reduction within the Breit-Pauli approximation. They demonstrated that
$d$ could be written $d=d_1 +d_2$, where $d_1$ is the electron
spin-rotational magnetic interaction and $d_2$ is a second-order term
in the electronic wave function.  They calculated values for $b$, $c$,
and $d_1$, estimated $d_2$, and calculated transition frequencies and
the constants $b$ and $c$ were used by Richardson, Jefferts, and
Dehmelt to make the first experimental estimate of
$d$~\cite{RicJefDeh68}.  Further improvements in calculations of $b$,
$c$, and $d_1$, and improved estimates of $d_2$ were presented by
Somerville~\cite{Som68,Som70} and by Luke~\cite{Luk69}. The
second-order constant $d_2$ was calculated accurately using a
variation-perturbation method by Kalaghan and Dalgarno~\cite{KalDal72}
and by McEachran, Veenstra, and Cohen~\cite{McEVeeCoh78}, who also
calculated $b$, $c$, and $d_1$ and accurate transition frequencies.

The calculated constants discussed so far were all  determined in
the Born-Oppenheimer approximation in which the electron is assumed to
move in the field generated by averaging over the electronic wave
function.  Nonadiabatic effects on the ground electronic state through
the excitation of other electronic states due to the coupling of
electronic and nuclear motion contribute significantly to the constant
$b$ and the effects of the reduced mass of the electron on the
constants have been treated~\cite{BabDal91,BabDal92}.

The nuclear spin-rotation term $f$ was introduced into the effective
spin Hamiltonian (\ref{ham}) on phenomenological grounds by
Jefferts~\cite{Jef69}, who measured the hfs transition frequencies for
the $v=4$--8 vibrational states.  For the $N=1$ rotational level for
each $v$, he obtained values of the constants by fitting to
(\ref{ham}). Varshalovich and Sannikov~\cite{VarSan93} obtained
different fits to Jefferts' data, notably, they found the value of $f$
to be negative.  The first (and to date the only) measurements of the
hfs of the $N=1$ level of $v=0$ state were carried out by Fu, Hessels,
and Lundeen~\cite{FuHesLun92} through spectroscopy of high angular
momentum Rydberg states of the hydrogen molecule~$\Hmol$ and by
fitting the measured Rydberg transition levels to an effective spin
Hamiltonian including~(\ref{ham}) they determined values for $b$, $c$,
$d$, and $f$.

A recent theoretical study of the nuclear spin-rotation interaction in
$\Hion$ found the sign of $f$ to be negative~\cite{Bab95}.  The
nuclear spin-rotation interaction can be written
\begin{equation}
\label{f-eq}
f(R) = f_1 (R) + f_2 (R) ,
\end{equation}
and the major contributions~\cite{Ram56} to the energy 
$hf$ are
from the interaction of each nuclear magnetic moment with the magnetic
field generated by the other rotating nucleus,
\begin{equation}
hf_1(R) = -\frac{4g_p\mu_N^2}{R^3}  ,
\end{equation}
and with the magnetic field generated by the orbiting
electron~\cite{Ram50d},
\begin{equation}
\label{f2}
hf_2(R) = -\frac{12g_p\mu_N^2}{\alpha^2R^2}  \sigma_{\rm hf} (R) ,
\end{equation}
where in atomic units $R$ is the internuclear distance, $\mu_N$ is the
nuclear magneton, and the dimensionless quantities $\alpha$, $g_p$,
and $\sigma_{\rm hf}$ are, respectively, the fs constant, the proton
$g$-factor defined previously, and the high-frequency component of the
magnetic shielding constant as defined in Eq.~(3) of~\cite{Bab95}.

In Table~\ref{compare} values of the hfs constants calculated as
in~\cite{BabDal92} and the nuclear spin-rotation constants calculated
as in~\cite{Bab95}, averaged over the vibrational wave functions, are
compared with constants measured~\cite{FuHesLun92} for the $N=1$ level
of the $v=0$ state and with constants determined~\cite{VarSan93} from
the measurements~\cite{Jef69} on the $N=1$ level of the vibrational
state $v=4$.  The agreement between theory and experiment is
impressive---the theoretical constants in Table~\ref{compare}
reproduce all measured transition frequencies to within 175~kHz.  The
formalism of~\cite{BabDal92} accounts for reduced mass effects on the
hfs constants but not fully for radiative and relativistic effects.

\section*{DEUTERON MOLECULAR ION}
The electric quadrupole moment $Q$ of the deuteron can serve as a
sensitive test of models of the neutron-proton nuclear
force~\cite{PreBha75}.  The most accurate determination
$Q=0.2860\pm0.0015 \times 10^{-26}\mbox{cm}^2$ was obtained
semiempirically from measurements~\cite{CodRam71} of the magnetic hfs
of the deuterium molecule $\Dmol$ through a theoretical value for the
electric field gradient at the nucleus due to the other constituent
charges of the molecule~\cite{ReiVai75,BisChe79}.  It should be
possible to measure $Q$ semiempirically using $\Dion$, which has the
theoretical advantage of a simpler electronic structure than $\Dmol$,
thereby facilitating accurate computation, but to date there have been
no measurements of the hfs of $\Dion$.  (An experiment that measured
hfs levels of $\HDion$ was reported, but the quadrupole effects were
not resolved~\cite{WinRufLam76}.)

For $\Dion$, the effective spin Hamiltonian is (\ref{ham}) plus an
additional nuclear quadrupole interaction~\cite{TowSch55-quadr}
proportional to $eqQ$, where $e$ is the proton charge and $q$ is the
electric field gradient.  Predictions of the hfs constants and
electric quadrupole coupling constant for the $N=1$ level of the $v=0$
state of $\Dion$ are given in Table~\ref{D-results}.  The constants
$b$, $c$, and $d$ have been calculated as in~\cite{BabDal92} with the
inclusion of nonadiabatic effects in $b$ and $c$, and $f$ has been
calculated using the theory of~\cite{Bab95}.  The term $eqQ$ was
estimated by calculating $q$ in the Born-Oppenheimer approximation and
using the semi-empirical value of $Q$ from $\Dmol$.  More detailed
calculations of the hfs constants, quadrupole coupling constants, and
transition frequencies will be reported~\cite{BabDal96}.

The quadrupole coupling constant is  about 57~kHz, see Table~\ref{D-results}.
The experimental precision achieved in the ion trap experiments on
$\Hion$ was better than 3~kHz~\cite{Jef69}, in the Rydberg $\Hmol$
experiment it was around 20~kHz~\cite{FuHesLun92}, and in a laser-rf
double-resonance ion beam study of $\mbox{N}_2{}^+$ it was better than
10~kHz~\cite{BerManKur91}. It would appear that the prospects are good
for measuring the deuteron quadrupole moment using $\Dion$.

\acknowledgements
The author is grateful to Y.~M. Cho and the International Organizing
Committee for support to attend the Inaugural Conference of the APCTP
and to S.~D. Oh, H.~W. Lee, and W. Jhe for their hospitality.
This work was supported in part by the National Science Foundation
through a grant for the Institute for Theoretical Atomic and Molecular
Physics at the Smithsonian Astrophysical Observatory and Harvard
University.


\begin{table}
\begin{center}
\caption{Comparison of $\Hion$  hfs constants from theory
with those obtained from experimental measurements of hfs levels, for
the $N=1$ rotational levels of the $v=0$ and $v=4$ vibrational states.
All values are given in MHz and numbers in parenthesis represent
errors quoted on the last digit.
\label{compare}}
\begin{tabular}{cllllc}
\multicolumn{1}{c}{ }   & \multicolumn{1}{c}{$b$} &
  \multicolumn{1}{c}{$c$} & \multicolumn{1}{c}{$d$} &
  \multicolumn{1}{c}{$f$} & \multicolumn{1}{c}{Ref.}  \\
\hline
Theory, $v=0$     &  880.163  & 128.482     & 42.421& $-0.042$ & 
   \protect\cite{BabDal92,Bab95}\\
Experiment, $v=0$ &  880.187(22)  & 128.259(26) & 42.348(29) & $-0.003(15)$ & 
   \protect\cite{FuHesLun92}\\
\multicolumn{6}{c}{ } \\
Theory, $v=4$     &  804.104 & 98.008 & 32.658 & $-0.036$ & 
   \protect\cite{BabDal92,Bab95}\\
Experiment, $v=4$ &  804.087(2) & 97.930(2) & 32.649(2) & $-0.034(2)$ & 
   \protect\cite{VarSan93}\\
\end{tabular}
\end{center}
\end{table}

\begin{table}
\begin{center}
\caption{Predictions of the hfs constants 
and the electric quadrupole coupling constant 
for the molecular ion $\Dion$
in the $N=1$ rotational state of the $v=0$ vibrational level.
All values are given in MHz.\label{D-results}}
\begin{tabular}{lllll}
   \multicolumn{1}{c}{$b$}  &
    \multicolumn{1}{c}{$c$} &
    \multicolumn{1}{c}{$d$} & \multicolumn{1}{c}{$f$} &
      \multicolumn{1}{c}{$eqQ$}\\
\hline
 135.739 & 19.944 & 21.461 & $-0.002$ & 0.057  \\
\end{tabular}
\end{center}
\end{table}

\end{document}